\documentclass[preprint]{elsarticle}

\usepackage{lineno,hyperref}
\usepackage{amsmath}
\usepackage[utf8x]{inputenc}
\usepackage{amssymb}
\usepackage{mathtools}
\usepackage{multirow}
\usepackage{booktabs}
\usepackage{array, threeparttable, booktabs}
\usepackage{filecontents}
\usepackage{url} 
\usepackage[labelfont=bf,skip=0pt, singlelinecheck=off,
    justification=justified]{caption}

\makeatletter
\renewcommand{\fnum@figure}{Fig. \thefigure.}
\makeatother
\usepackage{siunitx}
\usepackage{booktabs}
\usepackage[figuresright]{rotating}
\usepackage{blindtext}
\usepackage{pdflscape}
\usepackage{boxhandler}
\usepackage{epsfig,graphics}
\usepackage{etoolbox}
\usepackage{natbib}

\setcitestyle{authoryear,open={(},close={)}}

\hypersetup{colorlinks, urlcolor=blue, citecolor=blue}
\usepackage[usenames,dvipsnames,svgnames,table]{xcolor}
\makeatletter
\def\@footnotecolor{red}
\define@key{Hyp}{footnotecolor}{%
 \HyColor@HyperrefColor{#1}\@footnotecolor%
}
\patchcmd{\@footnotemark}{\hyper@linkstart{link}}{\hyper@linkstart{footnote}}{}{}
\makeatother
\hypersetup{footnotecolor=blue}

\makeatletter
\renewcommand\hyper@natlinkbreak[2]{#1} 
\makeatother

\makeatletter
\def\printFirstPageNotes{%
  \iflongmktitle
   \let\columnwidth=\textwidth\fi
  \ifx\@tnotes\@empty\else\@tnotes\fi
  \ifx\@nonumnotes\@empty\else\@nonumnotes\fi
  \ifx\@cornotes\@empty\else\@cornotes\fi
  \ifx\@elseads\@empty\relax\else
   \let\thefootnote\relax
   \footnotetext{\ifnum\theead=1\relax
      \textit{E-mail address:\space}\else
      \textit{E-mail addresses:\space}\fi
     \@elseads}\fi
  \ifx\@elsuads\@empty\relax\else
   \let\thefootnote\relax
   \footnotetext{\textit{URL:\space}%
     \@elsuads}\fi
  \ifx\@fnotes\@empty\else\@fnotes\fi
  \iflongmktitle\if@twocolumn
   \let\columnwidth=\Columnwidth\fi\fi
}
\makeatother

\modulolinenumbers[5]

\journal{Journal of Forecasting}

\begin{document}

\begin{frontmatter}

\title{Benchmark Dataset for Mid-Price Forecasting of Limit Order Book Data with Machine Learning Methods}

\author[mymainaddress]{Adamantios Ntakaris\corref{mycorrespondingauthor}}
\cortext[mycorrespondingauthor]{Corresponding author}
\ead{\string\href{mailto:adamantios.ntakaris@tuni.fi}{adamantios.ntakaris@tuni.fi}}

\author[mysecondaryaddress]{Martin Magris}

\author[mysecondaryaddress]{Juho Kanniainen}

\author[mymainaddress]{Moncef Gabbouj}

\author[mythirdaddress]{Alexandros Iosifidis}

\address[mymainaddress]{Laboratory of Signal Processing, Tampere University of Technology, Korkeakoulunkatu 1, Tampere, Finland}
\address[mysecondaryaddress]{Laboratory of Industrial and Information Management, Tampere University of Technology, Korkeakoulunkatu 8, Tampere, Finland }
\address[mythirdaddress]{Department of Engineering, Electrical and Computer Engineering, Aarhus University, Inge Lehmanns Gade 10, Aarhus, Denmark }

\begin{abstract}
Managing the prediction of metrics in high-frequency financial markets is a challenging task. An efficient way is by monitoring the dynamics of a limit order book to identify the information edge. This paper describes the first publicly available benchmark dataset of high-frequency limit order markets for mid-price prediction. We extracted normalized data representations of time series data for five stocks from the NASDAQ Nordic stock market for a time period of ten consecutive days, leading to a dataset of ≈4,000,000 time series samples in total. A day-based anchored cross-validation experimental protocol is also provided that can be used as a benchmark for comparing the performance of state-of-the-art methodologies. Performance of baseline approaches are also provided to facilitate experimental comparisons. We expect that such a large-scale dataset can serve as a testbed for devising novel solutions of expert systems for high-frequency limit order book data analysis.
\end{abstract}


\end{frontmatter}

\captionStyle{n}{l}
\section{Introduction}\label{SS:Intro}

Automated trading became a reality when the majority of exchanges adopted it globally. This environment is ideal for high-frequency traders. High-frequency trading (HFT) and a centralized matching engine, referred to as a limit order book (LOB), are the main drivers for generating big data \cite{seddon2017model}. In this paper, we describe a new order book dataset consisting of approximately four million events for ten consecutive trading days for five stocks. The data is derived from the ITCH feed provided by NASDAQ OMX Nordic and consists of the time-ordered sequences of messages that track and record all the events occurring in the specific market. It provides a complete market-wide history of ten trading days. Additionally, we define an experimental protocol to evaluate the performance of research methods in mid-price prediction.\footnote{Mid-price is the average of the best bid and best ask prices.}

Datasets, like the one presented here, come with challenges, including the selection of appropriate data transformation, normalization, description, and classification. This type of massive dataset requires a very good understanding of the available information that can be extracted for further processing. We follow the information edge, as has been recently presented by \cite{kercheval2015modelling}. The authors provide a detailed description of representations that can be used for a mid-price movement prediction metric. In light of this data representation, they apply non-linear classification based on support vector machines (SVM) in order to predict the movement of this metric. Such a supervised learning model exploits class labels\footnote{Labels are extracted from annotations provided by experts and represent the direction of the mid-price. Three different states are defined, i.e. upward, downward, and stationary movement.} for short- and long-term prediction. However, they train their model based on a very small (when compared to the size of the data that can be available for such applications) dataset of 4000 samples. This is due to the limitations of many non-linear kernel-based classification models related to their time and space complexity with respect to the training data size. On the other hand, \cite{sirignano2016deep} uses large amounts of data for non-linear classification based on a feedforward network. The author takes advantage of the local spatial structure\footnote{By local movement, the author means that the conditional movement of the future price (e.g. best ask price movement) depends, locally, on the current LOB state.} of the data for modelling the joint distribution of the LOB's state based on its current state.

Despite the major importance of publicly available datasets for advancing research in the HFT field, there are no detailed public available benchmark datasets for method evaluation purposes. In this paper, we describe the first publicly available dataset\footnote{The dataset can be downloaded from: \href{http://urn.fi/urn:nbn:fi:csc-kata20170601153214969115}{http://urn.fi/urn:nbn:fi:csc-kata20170601153214969115}} for an LOB-based HFT that has been collected in the hope of facilitating future research in the field. Based on \cite{kercheval2015modelling}, we provide time series representations approximately $4,000,000$  trading events and annotations for five classification problems. Baseline results of two widely used methods, i.e. linear and non-linear regression models, are also provided. In this way, we introduce this new problem for the expert systems community and provide a testbed for facilitating future research. We hope that attracting the interest of expert systems will lead to the rapid improvement of the performance achieved in the provided dataset, thus leading to much better state-of-the-art solutions to this important problem.
      
The dataset described in this paper can be useful for financial expert systems in two ways. First, it can be used to identify circumstances under which markets are stable, which is very important for liquidity providers (market makers) to make the spread. Consequently, such an intelligent system would be valuable as a framework that can increase liquidity provision. Secondly, analysis of the data can be used for model selection by speculative traders, who are trading based on their predictions on market movements. In future research, this paper can be employed to identify order book spoofing, i.e. situations where markets are exposed to manipulation by limit orders. In this case, spoofers could aim to move markets in certain directions by limit orders that are cancelled before they are filled. Therefore, this research is relevant not only for market makers and traders, but also for supervisors and regulators.

Therefore, the present work has the following contributions: 1) To the best of our knownledge this is the first publicly available LOB-ITCH dataset for machine learning experiments on the prediction of mid-price movements. 2) We provide baselines methods based on ridge regression and a new implementation of an RBF neural network based on k-means algorithm. 3) The paper provides information about the prediction of mid-price movements to market makers, traders, and regulators. This paper does not suggest any trading strategies and is reliant on purely machine learning metrics prediction. Overall, this work is an empirical exploration of the challenges that come with high-frequency trading and machine learning applications. 

The data from Nasdanq Helsinki Stock Exchange offers important benefits. In the United States the limit orders for a given asset are spread between several exchanges, causing fragmentation of liquidity. The fragmentation poses a problem for empirical research, because, as \cite{gould2013limit} point out, the ``differences between different trading platforms' matching rules and transaction costs complicate comparisons between different limit order books for the same asset.'' These issues related to fragmentation are not present with data obtained from less fragmented Nasdaq Nordic markets. Moreover, Helsinki Exchange is a pure limit order market, where the market makers have a limited role.

The rest of the paper is organized as follows. We provide a comprehensive literature review of the field in \hyperref[SS:MLHFT]{Section 2}. Dataset and experimental protocol descriptions are provided in \hyperref[SS:LOB]{Section 3}. Quantitative and qualitative comparisons of the new dataset, along with related data sources, are provided in \hyperref[SS:Existing]{Section 4}. In \hyperref[SS:Baselines]{Section 5}, we describe the engineering of our baselines. \hyperref[SS:Results]{Section 6} presents our empirical results and \hyperref[SS:Conclusion]{Section 7} concludes.

\section{Machine Learning for HFT and LOB}\label{SS:MLHFT}

The complex nature of HFT and LOB spaces is suitable for interdisciplinary research. In this section, we provide a comprehensive review of recent methods exploiting machine learning approaches. Regression models, neural networks, and several other methods have been proposed to make inferences of the stock market. Existing literature ranges metric prediction to optimal trading strategies identification. Research community has tried to tackle the challenges of prediction and data inference from different angles. While mid-price prediction can be considered a traditional time series prediction problem, there are several challenges that justify HFT as a unique problem.

\subsection{Regression Analysis}
\medskip

Regression models have been widely used for HFT and LOB prediction. \cite{zheng2012price} utilize logistic regression in order to predict the inter-trade price jump. \cite{alvim2010daily} use support vector regression (SVR) and partial least squares (PLS) for trading volume forecasting for ten Bovespa stocks. \cite{pai2005hybrid} use a hybrid model for stock price prediction. They combine an auto-regressive integrated moving average (ARIMA) model and an SVM classifier in order to model non-linearities of class structure in regression estimation models. \cite{liu2015behind} develop a multivariate linear model to explain short-term stock price movement where a bid-ask spread is used for classification purposes. \cite{detollenaere2017identifying} apply an adaptive least absolute shrinkage and selection operator (LASSO)\footnote{Adaptive weights are used for penalizing different coefficients in the $l_1$ penalty term.} for variable selection, which best explains the transaction cost of the split order. They apply an adjusted ordinal logistic method for classifying ex ante transaction costs into groups. \cite{cenesizoglu2014effects} work on a similar problem. They hold that the state of the limit order can be informative for the direction of future prices and try to prove their position by using an autoregressive model.

\cite{panayi2016designating} use generalized linear models (GLM) and generalized additive models for location, shape and scale (GAMLSS) models in order to relate the threshold exceedance duration (TED), which measures the length of time required for liquidity replenishment, to the state of the LOB. \cite{yu2006limit} tries to extract information from order information and order submission based on the ordered probit model.\footnote{The method is the generalization of a linear regression model when the dependent variable is discrete.} The author shows, in the case of Shanghai's stock market, that an LOB's information is affected by the trader's strategy, with different impacts on the bid and ask sides. \cite{amaya2015distilling} use panel regression\footnote{Panel regression models provide information on data characteristics individually, but also across both individuals over time.} for order imbalances and liquidity costs in LOBs so as to identify resilience in the market. Their findings show that such order imbalances cause liquidity issues that last for up to ten minutes. \cite{malik2014intraday} analyse the asymmetric intra-day patterns of LOBs. They apply regression with a power transformation on the notional volume weighted average price (NVWAP) curves in order to conclude that both sides of the market behave asymmetrically to market conditions.\footnote{Market conditions of an industry sector have an impact on sellers and buyers who are related to it. Factors to consider include the number of competitors in the sector. For example, if there is a surplus, new companies may find it difficult to enter the market and remain in business.} In the same direction, \cite{ranaldo2004order} examines the relationship between trading activity and the order flow dynamics in LOBs, where the empirical investigation is based on a probit model. \cite{cao2009information} examine the depth of different levels of an order book by using an autoregressive (AR) model of order 5 (the AR(5) framework). They find that levels beyond the best bid and best ask prices provide moderate information regarding the true value of an asset. Finally, \cite{creamer2012model} suggests that the LogitBoost algorithm is ideal for selecting the right combination of technical indicators.\footnote{Technical indicators are mainly used for short-term price movement predictions. They are formulas based on historical data.}

\subsection{Neural Networks}
\medskip

HFT is mainly a scalping\footnote{Scalping is a type of trading strategy according to which the trader tries to make a profit for small changes in a stock.} strategy according to which the chaotic nature of the data creates the proper framework for the application of neural networks. \cite{levendovszky2012prediction} propose a multi-layer feedforward neural network for predicting the price of EUR/USD pair, trained by using the backpropagation algorithm. \cite{sirignano2016deep} proposes a new method for training deep neural networks that tries to model the joint distribution of the bid and ask depth, where a focal point is the spatial nature\footnote{The spatial nature of this type of neural network and its gradient can be evaluated at far fewer grid points. This makes the model less computationally expensive. Furthermore, the suggested architecture can model the entire distribution in the $R^d$ space.} of LOB levels. \cite{bogoev2016empirical} propose the use of a single-hidden layer feedforward neural (SLFN) network for the detection of quote stuffing and momentum ignition.  \cite{dixon2016high} uses a recurrent neural network (RNN) for mid-price predictions of T-bond\footnote{Treasury bond (T-bond) is a long-term fixed interest rate debt security issued by the federal government.} and ES futures\footnote{E-mini S\&P 500 (ES futures) are electronically traded futures contracts whose value is one-fifth that size of standard S\&P futures.} based on ultra-high-frequency data. \cite{rehman2014foreign} apply recurrent cartesian genetic programming evolved artificial neural network (RCGPANN) for predicting five currency rates against the Australian dollar. \cite{galeshchuk2016neural} suggests that a multi-layer perceptron (MLP) architecture, with three hidden layers, is suitable for exchange rate prediction. \cite{majhi2009development} use the Functional Link Artificial Neural Network (FLANN) in order to predict price movements in the DJIA\footnote{The Dow Jones Industrial Average (DJIA) is the price-weighted average of the 30 largest, publicly-owned U.S. companies.} and S\&P500\footnote{S\&P500 is the index that provides a summary of the overall market by tracking some of the 500 top stocks in U.S. stock market.} stock indices.

Deep belief networks employed by \cite{sharang2015using} to design a medium-frequency portfolio trading strategy. \cite{hallgren2016testing} use continuous time bayesian networks (CTBNs) for causality detection. They apply their model on tick-by-tick high frequency foreign exchange (FX) data EUR/USD by using a Skellam process.\footnote{A Skellam process is defined as $S(t) = N^{(1)}(t)-N^{(2}(t), \ t\geqslant0$ where $N^{(1)}(t)$ and $N^{(2)}(t)$ are two independent homogeneous Poisson processes.} \cite{sandoval2015computational} create a profitable trading strategy by combining hierarchical hidden Markov models (HHMM), where they consider wavelet-based LOB information filtering. In their work, they also consider a two-layer feedforward neural network in order to classify the upcoming states. They nevertheless report limitations in the neural network in terms of the volume of the input data.

\subsection{Maximum Margin and Reinforcement Learning}
\medskip

\cite{palguna2016mid} use nonparametric methods on features derived from LOB, which are incorporated into order execution strategies for mid-price prediction. In the same direction, \cite{kercheval2015modelling} employ a multi-class SVM for mid-price and price spread crossing prediction.  \cite{han2015machine} base their research on \cite{kercheval2015modelling} by using multi-class SVM for mid-price movement prediction. More precisely, they compare multi-class SVM (exploring linear and RBF kernels) to decision trees using bagging for variance reduction.

\cite{kim2001input} uses input/output hidden Markov models (IOHMMs) and reinforcement learning (RL) in order to identify the order flow distribution and market-making strategies, respectively. \cite{yang2015gaussian} apply apprenticeship learning\footnote{Motivation for
apprenticeship learning is to use IRL techniques to learn the reward function and then use this function in order to define a Markov decision problem (MDP).} methods, like linear inverse reinforcement learning (LIRL) and Gaussian process IRL (GPIRL), to recognize traders or algorithmic trades based on the observed limit orders. \cite{chan2001electronic} use RL for market-making strategies, where experiments based on a Monte Carlo simulation and a state-action-reward-state-action (SARSA) algorithm test the efficacy of their policy. In the same vein, \cite{kearns2013machine} implement RL for trade execution optimization in lit and dark pools. Especially in the case of dark pools, they apply a censored exploration algorithm to the problem of smart order routing (SOR). \cite{yang2012behavior} examine an IRL algorithm for the separation of HFT strategies from other algorithmic trading activities. They also apply the same algorithm to the identification of manipulative HFT strategies (i.e. spoofing). \cite{felker2014distance} predict changes in the price of quotes from several exchanges. They apply feature-weighted Euclidean distance to the centroid of a training cluster. They calculate this type of distance to the centroid of a training cluster where feature selection is taken into consideration because several exchanges are included in their model.

\subsection{Additional Methods for HFT and LOB}
\medskip

HFT and LOB research activity also covers topics like the optimal submission strategies of bid and ask orders with a focus on the inventory risk that stems from an asset's value uncertainty, as in the work of \cite{avellaneda2008high}. \cite{chang2015inferring} models the dynamics of LOB by using a Bayesian inference of the Markov chain model class, tested on high-frequency data. \cite{an2017short} suggest a new stochastic model which is based on independent compound Poisson processes of the order flow. \cite{talebi2014multi} try to predict trends in the FX market by employing a multivariate Gaussian classifier (MGC) combined with Bayesian voting. \cite{fletcher2010multiple} examine trading opportunities for the EUR/USD where the price movement is based on multiple kernel learning (MKL). More specifically, the authors utilize SimpleMKL, and the more recent LPBoostMKL, methods for training a multi-class SVM. \cite{christensen2013prediction} develop a classification method based on Gaussian kernel in order to identify iceberg\footnote{Iceberg order is the conditional request made to the broker to sell or buy a larger quantity of the stock, but in smaller predefined quantities.} orders for GLOBEX.

\cite{maglaras2015optimal} consider the LOB as a multi-class queueing system in order to solve the problem placement of limit and market order placements. \cite{mankad2013discovering} apply a static plaid clustering technique to synthetic data in order to classify the different types of trades. \cite{aramonte2013assessing} show that the information asymmetry in a high-frequency environment is crucial.

\cite{vella2016improving} use higher order fuzzy systems (i.e. an adaptive neuro-fuzzy inference system) by introducing T2 fuzzy sets where the goal is to reduce microstructure noise in the HFT sphere.  \cite{abernethy2013adaptive} apply market-maker strategies based on low regret algorithms for the stock market. \cite{almgren2006bayesian} explain price momentum by modelling Brownian motion with a drift whose distribution is updated based on Bayesian inference. \cite{naes2006order} show that the order book slope measures the elasticity of supplied quantity as a function of asset prices related to volatility, trading activity, and an asset's dispersion beliefs.

\section{The LOB Dataset}\label{SS:LOB}

In this section, we describe in detail our dataset collected in order to facilitate future research in LOB-based HFT. We start by providing a detailed description of the data in \hyperref[SS:DataDescription]{Section} \ref{SS:DataDescription}. Data processing steps are followed in order to extract message books and LOBs,
 as described in \hyperref[SS:MessageLimitOrderBooks]{Section} \ref{SS:MessageLimitOrderBooks}.

\subsection{Data Description}\label{SS:DataDescription}
\medskip

\begin{figure}[ht!]
\centering
\includegraphics[scale=0.85]{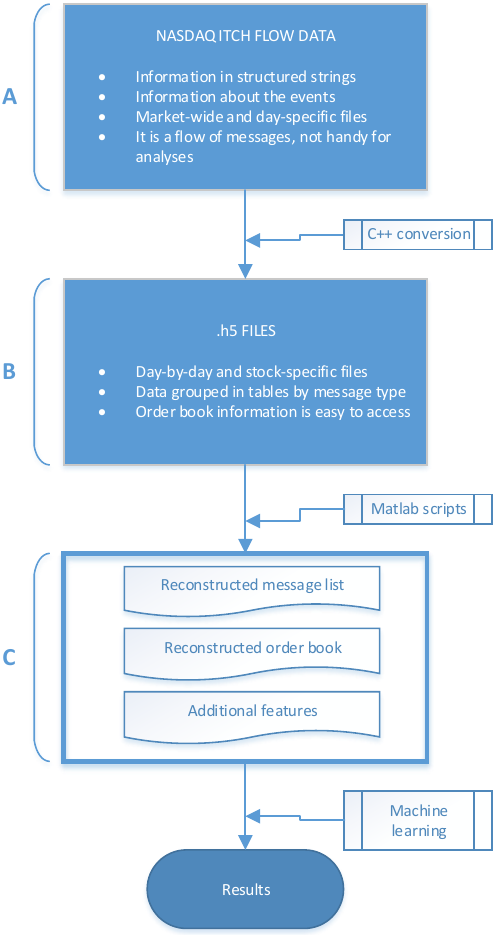}
\captionsetup{justification=centering}
\caption{Data processing flow}
\label{fig:data_flow}
\end{figure}

Extracting information from the ITCH flow, and without relying on third-party data providers, we analyse stocks from different industry sectors for ten full days of ultra-high-frequency intra-day data. The data provides information regarding trades against hidden orders. Coherently, the non-displayable hidden portions of the total volume of a so-called iceberg order are not accessible from the data. Our ITCH feed data is day-specific and market-wide, which means that we deal with one file per day with data over all the securities. Information (block A in \hyperref[fig:data_flow]{Fig.} \ref{fig:data_flow}) regarding (i) messages for order submissions, (ii) trades, and (iii) cancellations, is included. For each order, its type (buy/sell), price, quantity, and exact time stamp on a millisecond basis is available. In addition, (iv) administrative messages (i.e. trading halts or basic security data), (v) event controls (i.e. start and ending of trading days, states of market segments), and (vi) net order imbalance indicators are also included.

The next step is the development and implementation of a C++ converter to extract all the information relevant to a given security. We perform the same process for five stocks traded on the NASDAQ OMX Nordic at the Helsinki exchange from 1 June 2010 to 14 June 2010\footnote{There have been about 23,000 active order books, the vast majority of which are very illiquid, show sporadic activity, and correspond to little and noisy data.}. This data is stored in a Linux cluster. Information related to the five stocks is illustrated in \hyperref[tab:stock_info]{Table} \ref{tab:stock_info}. The selected stocks\footnote{The choice is driven by the necessity of having a sufficient amount of data for training (this excludes illiquid stocks) while covering different industry sectors. These five selected stocks (see \hyperref[tab:stock_info]{Table} \ref{tab:stock_info}), which aggregate input message list and order book data for feature extraction, are about 4GB; RTRKS was suspended from trading and delisted from the Helsinki exchange on 20 Nov 2014.} are traded in one exchange (Helsinki) only. By choosing only one stock market exchange, the trader has the advantage of avoiding issues associated with fragmented markets. In the case of fragmented markets, the limit orders for a given asset are spread between several exchanges, posing problems from empirical data analysis \cite{o2011market}.

\begin{table}[h!]
\centering
\captionsetup{width=.963\textwidth}
\caption[caption]{\\\hspace{\textwidth} Stocks used in the analysis}
\scalebox{0.8}{
\begin{tabular}{l l l l l}
\hline
Id & ISIN Code & Company & Sector & Industry\\
\hline
KESBV & FI0009000202 & Kesko Oyj 				& Consumer Defensive 		& Grocery Stores\\
OUT1V & FI0009002422 & Outokumpu Oyj 			& Basic Materials 			& Steel\\
SAMPO & FI0009003305 & Sampo Oyj 				& Financial Services 		& Insurance\\
RTRKS & FI0009003552 & Rautaruukki Oyj 			& Basic Materials 			& Steel\\
WRT1V & FI0009000727 & W\"{a}rtsil\"{a} Oyj 	& Industrials 				& Diversified Industrials\\
\hline
\end{tabular}}
\label{tab:stock_info}
\end{table}

The Helsinki Stock Exchange, operated by NASDAQ Nordic, is a pure electronic limit order market. The ITCH feed keeps a record of all the events, including those that take place outside active trading hours. At the Helsinki exchange, the trading period goes from 10:00 to 18:25 (local time,  UTC/GMT +2 hours). However, in the ITCH feed, we observe several records outside those trading hours. In particular, we consider the regulated auction period before 10:00, which is used to set the opening price of the day (the so-called pre-opening period) before trading begins. This is a structurally different mechanism following different rules with respect to the order book flow during trading hours. Similarly, another structural break in the order book's dynamics is due to the different regulations that are in force between 18:25 and 18:30 (the so-called post-opening period). As a result, we retain exclusively the events occurring between 10:30 and 18:00. More information related to the above-mentioned issues can be found in \cite{siikanen2016limit} and \cite{siikanen2016drives}. Here, the order book is expected to have comparable dynamics with no biases or exceptions caused by its proximity to the market opening and closing times.

\subsection{Limit Order and Message Books}\label{SS:MessageLimitOrderBooks}
\medskip

Message and limit order books are processed for each of the 10 days for the five stocks. More specifically, there are two types of messages that are particularly relevant here: (i) \lq\lq add order messages\rq\rq, corresponding to order submissions, and (ii) \lq\lq modify order messages\rq\rq, corresponding to updates on the status of existing orders through order cancellations and order executions. Example message\footnote{A sample from FI0009002422 on 01 June 2010.} and limit order\footnote{A sample from FI0009002422 on 01 June 2010.} books are illustrated in \hyperref[tab:mb_example]{Table} \ref{tab:mb_example} and \hyperref[tab:ob_example]{Table} \ref{tab:ob_example}, respectively.
\begin{table}[h!]
\centering
\captionsetup{width=.73\textwidth}
\caption[caption]{\\\hspace{\textwidth} Message list example}
\scalebox{0.8}{
\begin{tabular}{llllll}
\hline
Timestamp     & Id 	 & Price  & Quantity   & Event & Side \\
\hline
1275386347944 & 6505727 & 126200 & 400      & Cancellation & Ask  \\
1275386347981 & 6505741 & 126500 & 300      & Submission   & Ask  \\
1275386347981 & 6505741 & 126500 & 300      & Cancellation & Ask  \\
1275386348070 & 6511439 & 126100 & 17       & Execution    & Bid  \\
1275386348070 & 6511439 & 126100 & 17       & Submission   & Bid  \\
1275386348101 & 6511469 & 126600 & 300      & Cancellation & Ask  \\
\hline
\end{tabular}}
\label{tab:mb_example}
\end{table}

LOB is a centralized trading method that is incorporated by the majority of exchanges globally. It aggregates the limit orders of both sides (i.e. the ask and bid sides) of the stock market (e.g. the Nordic stock market). LOB matches every new event type according to several characteristics. Event types and LOB characteristics describe the current state of this matching engine. Event types can be executions, order submissions, and order cancellations. Characteristics of LOB are the resolution parameters \cite{gould2013limit}, which are the tick size $\pi$ (i.e. the smallest permissible price between different orders), and the lot size $\sigma$ (i.e. the smallest amount of a stock that can be traded and is defined as $\{ \textit{k}\sigma | \textit{k} = 1,2,...\}$). Order inflow and resolution parameters will formulate the dynamics of the LOB, whose current state will be identified by the state variable of four elements  $(s_t^b, q_t^b, s_t^a, q_t^a  ), {t\geq0}$, where $s_t^{b}$ ($s_t^{b}$) is the best bid (ask) price and $q_t^b$ ($q_t^a$) is the size of the best bid (ask) level at time t.

In our data, timestamps are expressed in milliseconds based on 1 Jan 1970 format and shifted by three hours with respect to Eastern European Time (in the data, the trading day goes from 7:00 to 15:25). ITHC feed prices are recorded up to 4 decimal and, in our data, the decimal point is removed by multiplying the price by 10,000 where currency is in Euro for the Helsinki exchange. The tick size, defined as the smallest possible gap between the ask and bid prices, is one cent. Similarly, orders' quantities are constrained to integers greater than one. 

\begin{table}[h!]
\centering
\captionsetup{width=.99\textwidth}
\caption[caption]{\\\hspace{\textwidth} Order book example}
\scalebox{0.6}{
\begin{tabular}{llllllllllll}
\hline
& & &\multicolumn{4}{c}{Level 1}&\multicolumn{4}{c}{Level 2} & ...\\
 \cmidrule(l){4-7}  \cmidrule(l){8-11} \cmidrule(l){12-11}
& & &\multicolumn{2}{c}{Ask} & \multicolumn{2}{c}{Bid} & \multicolumn{2}{c}{Ask} & \multicolumn{2}{c}{Bid} &  \\
\cmidrule(l){1-3} \cmidrule(l){4-5} \cmidrule(l){6-7} \cmidrule(l){8-9} \cmidrule(l){10-11}
Timestamp     & Mid-price & Spread & Price & Quantity & Price & Quantity& Price & Quantity& Price & Quantity & \\
\cmidrule(l){1-3}  \cmidrule(l){4-12}
1275386347944 & 126200     & 200    & 126300    & 300       & 126100    & 17        & 126400    & 4765      & 126000    & 2800      & ...   \\
1275386347981 & 126200     & 200    & 126300    & 300       & 126100    & 17        & 126400    & 4765      & 126000    & 2800      & ...   \\
1275386347981 & 126200     & 200    & 126300    & 300       & 126100    & 17        & 126400    & 4765      & 126000    & 2800      & ...   \\
1275386348070 & 126050     & 100    & 126100    & 291       & 126000    & 2800      & 126200    & 300       & 125900    & 1120      & ...   \\
1275386348070 & 126050     & 100    & 126100    & 291       & 126000    & 2800      & 126200    & 300       & 125900    & 1120      & ...   \\
1275386348101 & 126050     & 100    & 126100    & 291       & 126000    & 2800      & 126200    & 300       & 125900    & 1120      & ...   \\
\hline
\end{tabular}}
\label{tab:ob_example}
\end{table}

\subsection{Data Availability and Distribution}
\medskip

In compliance with NASDAQ OMX agreements, the normalized feature dataset is made available to the research community.\footnote{We thank Ms. Sonja Salminen at NASDAQ for her support and help.} The open-access version of our data has been normalized in order to prevent reconstruction of the original NASDAQ data. 

\subsection{Experimental Protocol}\label{SS:ExperimentalProtocol}
\medskip

\begin{figure}[h!]
\centering
\includegraphics[scale=0.33]{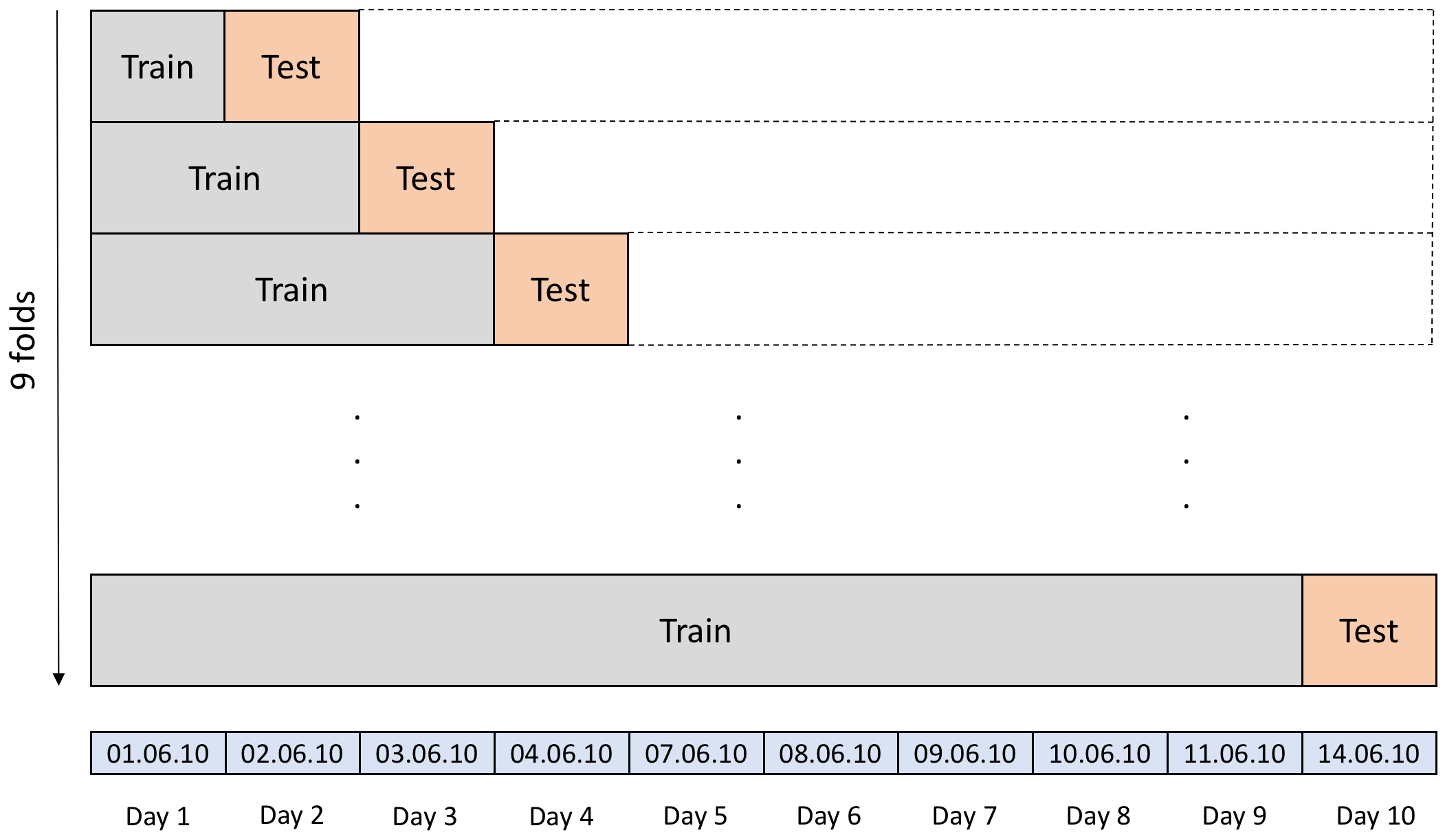}
\captionsetup{justification=centering}
\caption{Experimental Setup Framework}
\label{fig:Exp_Setup_3}
\end{figure}

In order to make our dataset a benchmark that can be used for the evaluation of HTF methods based on LOB information, the data is accompanied by the following experimental protocol. We develop a day-based prediction framework following an anchored forward cross-validation format. More specifically, the training set is increases by one day in each fold and stops after $n-1$ days (i.e. after 9 days in our case where $n=10$). On each fold, the test set corresponds to one day of data, which moves in a rolling window format. The experimental setup is illustrated in \hyperref[fig:Exp_Setup_3]{Fig.} \ref{fig:Exp_Setup_3}. Performance is measured by calculating the mean accuracy, recall, precision, and F1 score over all folds, as well as the corresponding standard deviation. We measure our results based on these metrics, which are defined as follows: 

\begin{equation}
Accuracy = \frac{TP + TN}{TP + TN + FP + FN}
\end{equation}

\begin{equation}
Precision = \frac{TP}{TP + FP}
\end{equation}

\begin{equation}
Recall = \frac{TP}{TP + FN}
\end{equation}

\begin{equation}
F1 = 2\times \frac{Precision \times Recall}{Precision + Recall}
\end{equation}

\noindent where TP and TF represents the true positives and true negatives, respectively, of the mid-price prediction label compared with the ground truth, where FP and FN represents the false positives and false negatives, respectively. From among the above metrics, we focus on the F1 score performance. The main reason that we focus on F1 score is based on its ability to only be affected in one direction of skew distributions, in the case of unbalanced classes like ours. On the contary, accuracy cannot differentiate between the number of correct labels (i.e. related to mid-price movement direction prediction) of different classes where the other three metrics can separate the correct labels among different classes with F1 being the harmonic mean of Precision and Recall.

We follow an event-based inflow, as used in \cite{li2016empirical}. This is due to the fact that events (i.e. orders, executions, and cancellations) do not follow a uniform inflow rate. Time intervals between two consecutive events can vary from milliseconds to several minutes of difference. Event-based data representation avoids issues related to such big differences in data flow. As a result, each of our representations is a vector that contains information for 10 consecutive events. Event-based data description leads to a dataset of approximately half a million representations (i.e. 394,337 representations). We represent these events using the $144$-dimensional representation proposed recently by \cite{kercheval2015modelling}, formed by three types of features: a) the raw data of a 10-level limit order containing price and volume values for bid and ask orders, b) features describing the state of the LOB, exploiting past information, and c) features describing the information edge in the raw data by taking time into account. Derivations of time, stock price, and volume are calculated for short and long-term projections. More specifically, types in features $u_7, u_8$, and $u_9$  are: \textit{trades, orders, cancellations, deletion, execution of a visible limit order}, and \textit{execution of a hidden limit order }. Expressions used for calculating these features are provided in \hyperref[tbl:Features]{Table} \ref{tbl:Features}. One limitation of the adopted features is the lack of information related to order flow (i.e. the sequence of order book messages). However, as can be seen in the Results Section 6,, the baselines achieve relatively good performance and therefore we leave the introduction of extra features that can enhance performance to future research.

We provide three sets of data, each created by following a different data normalization strategy, i.e. z-score, min-max, and decimal precision normalization, for every i datasample. Z-score, in particular, is the normalization process through which we subtract the mean from our input data for each feature separately and divide by the standard deviation of the given sample:
\begin{equation}\label{Eq:znorm}
\textbf{x}_{i}^{(Z_{score})} = \frac{\textbf{x}_i - \frac{1}{N} \sum\limits_{j=1}^{N}\textbf{x}_j}{\sqrt{\frac{1}{N}\sum\limits_{j=1}^{N}(\textbf{x}_j-\bar{\textbf{x}})^2}},
\end{equation}
where $\bar{\textbf{x}}$ denotes the mean vector, as appears in \hyperref[Eq:znorm]{Eq.} \ref{Eq:znorm}. 
\noindent
On the other hand, min-max scaling, as described by:
\begin{equation}
\textbf{x}_{i}^{(MM)} = \frac{\textbf{x}_i-\textbf{x}_{min}}{\textbf{x}_{max}-\textbf{x}_{min}},
\end{equation}
is the process of subtracting the minimum value from each feature and dividing it by the difference between the maximum and minimum value of that feature sample. The third scaling setup is the decimal precision approach. This normalization method is based on moving the decimal points of each of the feature values. Calculations follow the absolute value of each feature sample:

\begin{equation}
\textbf{x}_{i}^{(DP)} = \frac{\textbf{x}_i}{10^k},
\end{equation}
where k is the integer that will give us the maximum value for $|\textbf{x}_{DP}|<1$.

\begin{table}[h!]
\centering
\captionsetup{width=.989\textwidth}
\caption[caption]{\\\hspace{\textwidth} Feature Sets}\label{tbl:Features}
\scalebox{0.62}{
\begin{tabular}{rll}
\hline
Feature Set & Description & Details   \\
\hline
Basic 				&	$ u_1 = \{ P_i^{ask}, V_i^{ask}, P_i^{bid}, V_i^{bid}\}_{i=1}^n$  											&	10(=n)-level LOB Data\\
\\
Time-Insensitive		& 	$ u_2 = \{(P_i^{ask}-P_i^{bid}), (P_i^{ask}+P_i^{bid})/2 \}_{i=1}^n $										&	Spread \& Mid-Price\\
					& 	$ u_3 = \{P_n^{ask}-P_1^{ask}, P_1^{bid}-P_n^{bid}, |P_{i+1}^{ask}-P_i^{ask}|, |P_{i+1}^{bid}-P_i^{bid}| \}_{i+1}^n$	&	Price Differences\\
					& 	$ u_4 = \Big\{ \frac{1}{n}\sum\limits_{i=1}^{n}P_i^{ask},  \frac{1}{n}\sum\limits_{i=1}^{n}P_i^{bid},  \frac{1}{n}\sum\limits_{i=1}^{n}V_i^{ask},  \frac{1}{n}\sum\limits_{i=1}^{n}V_i^{bid}\Big\}$		&	Price \& Volume Means	\\

					& 	$ u_5 = \Big\{ \sum\limits_{i=1}^{n}(P_i^{ask} - P_i^{bid}), \sum\limits_{i=1}^{n}(V_i^{ask} - V_i^{bid})  \Big\}$			&	Accumulated Differences			\\
\\
					
Time-Sensitive  		&	$u_6 = \Big\{dP_i^{ask}/{dt}, dP_i^{bid}/{dt}, dV_i^{ask}/{dt}, dV_i^{bid}/{dt} \Big\}_{i=1}^n $		&	Price \& Volume Derivation			\\
					&	$u_7 = \Big\{ \lambda_{\Delta t}^{1}, \lambda_{\Delta t}^{2}, \lambda_{\Delta t}^{3}, \lambda_{\Delta t}^{4}, \lambda_{\Delta t}^{5}, \lambda_{\Delta t}^{6}  \Big\}$		&	Average Intensity per Type			\\
					&	$u_8 = \Big\{ \textbf{1}_{\lambda_{\Delta_t}^{1}>\lambda_{\Delta_T}^{1}}, \textbf{1}_{\lambda_{\Delta_t}^{2}>\lambda_{\Delta_T}^{2}}, \textbf{1}_{\lambda_{\Delta_t}^{3}>\lambda_{\Delta_T}^{3}}, \textbf{1}_{\lambda_{\Delta_t}^{4}>\lambda_{\Delta_T}^{4}}, \textbf{1}_{\lambda_{\Delta_t}^{5}>\lambda_{\Delta_T}^{5}}, \textbf{1}_{\lambda_{\Delta_t}^{6}>\lambda_{\Delta_T}^{6}} \Big\}$  	&	Relative Intensity Comparison 			\\
					&	$u_9 = \{d\lambda^{1}/dt, d\lambda^{2}/dt, d\lambda^{3}/dt, d\lambda^{4}/dt, d\lambda^{5}/dt, d\lambda^{6}/dt  \}$  	&	Limit Activity Accelaration 			\\					
\hline
\end{tabular}}

\end{table}

\noindent
Having defined the event representations, we use five different projection horizons for our labels. Each of these horizons portrays a different future projection interval of the mid-price movement (i.e. upward, downward, and stationary mid-price movement). More specifically, we extract labels based on short-term and long-term, event -based, relative changes for the next 1, 2, 3, 5, and 10 events for our representations dataset. 

Our labels describe the percentage change of the mid-price, which is calculated as follows:
\begin{equation}
l_{i}^{(j)} = \frac{\frac{1}{k}\sum\limits_{j=i+1}^{i+k}m_{j}-m_{i}}{m_{i}},
\end{equation} 
where $m_{j}$ is the future mid-prices ($k$ = 1, 2, 3, 5, or 10 next events in our representations) and $m_{i}$ is the current mid-price. The extracted labels are based on a threshold for the percentage change of 0.002. For percentage changes equal to or greater than 0.002, we use label 1. For percentage change that varies from -0.00199 to 0.00199, we use label 2, and, for percentage change smaller or equal to -0.002, we use label 3.

\section{Existing Datasets Described in the Literature}\label{SS:Existing}

In this section, we list existing HFT datasets described in the literature and provide qualitative and quantitative comparisons to our dataset. The following works mainly focus on datasets that are related to machine learning methods.

There are mainly three sources of data from which a high-frequency trader can choose. The first option is the use of publicly available data (e.g. (1) Dukascopy and (2) truefx), where no prior agreement is required for data acquisition. The second option is publicly available data upon request for academic purposes, which can be found in (3) \cite{brogaard2014high}, (4) \cite{hasbrouck2013low}, (5) \cite{de2007hide}, \cite{detollenaere2017identifying} and \cite{carrion2013very}. Finally, the third and most common option is data through platforms requiring a subscription fee, like those in (6) \cite{kercheval2015modelling}, \cite{li2016empirical}, and (7) \cite{sirignano2016deep}. Existing data sources and characteristics are listed in \hyperref[tbl:OtheDatasets]{Table} \ref{tbl:OtheDatasets}.
\begin{table}[h!]
\centering
\captionsetup{width=.977\textwidth} 
\caption[caption]{\\\hspace{\textwidth} HFT Dataset Examples}\label{tbl:OtheDatasets}
\scalebox{0.558888}{
\begin{tabular}{crcccccccc}
\hline
&Dataset & Public Avl. & Unit Time & Period & Asset Class / Num. of Stocks  & Size& Annotations \\
\hline
1 	&Dukascopy		&	\checkmark  	& 	ms	&      up-to-date 			& 	various				&	$\approx$ 20,000 events/day	&	\ding{53}		\\
2 	&truefx			& 	\checkmark 	&	ms	&	up-to-date			&	15 FX pairs			&	$\approx$ 300,000 events/day	&	\ding{53}		\\
3 	&NASDAQ			& 	AuR			&	ms	&	2008-09				&	Equity / 120			&	-						&	\ding{53}		\\
4 	&NASDAQ			& 	AuR			&	ms	&	10/07 \& 06/08			&	Equity / 500 			&	$\approx$ 55,000 events/day	&	\ding{53}		\\
5	&NASDAQ 		&	\ding{53}		&	ms	&	-					&	Equity / 5  			&	2,000 data points			&	\ding{53}		\\
6	&Euronext 		&	AuR			&	-	&	-					&	Several Products   		&	-						&	\ding{53}		\\
7	&NASDAQ			&	\ding{53}		&	ns  	&	01/14-08/15			&	Equity / 489			&	50 TB					&	\ding{53}		\\
8	& Our - NASDAQ	&	\checkmark  	&	ms	&	01-14/06/10			&	Equity / 5 				&	4 M samples				&	\checkmark	\\
\hline
\end{tabular}}
\end{table}

In particular, the datasets are in millisecond resolution, except for number six in the table. Access to various asset classes including FX, commodities, indices, and stocks is also provided. To the best of our knowledge, there is no available literature based on this type of dataset for equities.  Another source of free tick-by-tick historical data is the  truefx.com site, but the site provides data only for the FX market for several pairs of currencies in a millisecond resolution. The data contains information regarding timestamps (in millisecond resolution) and bid and ask prices. Each of these .csv files contains approximately 200,000 events per day. This type of data is used in a mean-reverting jump-diffusion model, as presented in \cite{PavarichEstimation2016}. 

There is a second category of datasets available upon request (AuR), as seen in \cite{hasbrouck2013low}. In this paper, the authors use the NASDAQ OMX ITCH for two periods: October 2007 and June 2008. For that period, they run samples of ten-minutes intervals for each day where they set a cut-off mechanism for available messages per period.\footnote{The authors provide a threshold, which is based on 250 events per 10-minute sample interval.} The main disadvantage of uniformly sampling HFT data is that the trader loses vital information. Events come randomly, with inactive periods varying from a few milliseconds to several minutes or hours. In our work, we overcome this challenge by considering the information based on event inflow, rather than equal time sampling. Another example of data that is available only for academic purposes is \cite{brogaard2014high}. The dataset contains information regarding timestamps, price, and buy-sell side prices but no other details related to daily events or feature vectors. In \cite{hasbrouck2013low}, the authors provide a detailed description of their NASDAQ OMX ITCH data, which is not directly accessible for testing and comparison with their baselines. They use this data to applying low-latency strategies based on measures that capture links between submissions, cancellations, and executions. Authors in \cite{de2007hide} and \cite{detollenaere2017identifying} use similar datasets from Euronext for limit order book construction. They specify that their dataset is available upon request from the provider. What is more, the data provider supplies with details regarding the LOB construction by the user. Our work fills that gap since our dataset provides the full limit order book depth and it is ready for use and comparison to our baselines.

The last category of dataset has dissemination restrictions. An example is the paper by  \cite{kercheval2015modelling}, where the authors are trying to predict the mid-price movement by using machine learning (i.e. SVM). They train their model with a very small number of samples (i.e. 4000 samples). The HFT activity can produce a huge volume of trading events daily, like our database does with 100,000 daily events for only one stock. Moreover, the datasets in \cite{kercheval2015modelling} and \cite{sirignano2016deep} are not publicly available, which makes comparison with other methods impossible. In the same direction, we also add works such as \cite{hasbrouck2009trading}, \cite{kalay2004measuring}, and \cite{kalay2002continuous} which utilize TAQ and Tel-Aviv stock exchange datasets (not for machine learning methods), and require subscription.

\section{Baselines}\label{SS:Baselines}

In order to provide performance baselines for our new dataset of HFT with LOB data, we conducted experiments with two regression models using the data representations described in \hyperref[SS:ExperimentalProtocol]{Section} \ref{SS:ExperimentalProtocol}. Details on the models used are provided in \hyperref[SS:RR]{Section} \ref{SS:RR} and \hyperref[SS:PVM]{Section} \ref{SS:PVM}. The baseline performances are provided in \hyperref[SS:Results]{Section} \ref{SS:Results}.

\subsection{Ridge Regression (RR)}\label{SS:RR}
\medskip

Ridge regression defines a linear mapping, expressed by the matrix $\mathbf{W}\in \mathbb{R}^{D \times C}$, that optimally maps a set of vectors $\mathbf{x}_i \in \mathbb{R}^{D}, \:i=1,\dots,N$ to another set of vectors (noted as target vectors) $\mathbf{t}_i \in \mathbb{R}^{C}, \:i=1,\dots,N$, by optimizing the following criterion:
\begin{equation}
\mathbf{W}^* = \underset{\mathbf{W}}{
\arg min} \:\: \sum_{i=1}^N \| \mathbf{W}^T \mathbf{x}_i - \mathbf{t}_i \|^2_2 + \lambda \|\mathbf{W}\|^2_F,
\end{equation}
or using a matrix notation:
\begin{equation}\label{Eq:RRcriterion}
\mathbf{W}^* = \underset{\mathbf{W}}{\arg min} \:\: \| \mathbf{W}^T \mathbf{X} - \mathbf{T} \|^2_F + \lambda \|\mathbf{W}\|^2_F.
\end{equation}
In the above, $\mathbf{X} = [\mathbf{x}_i,\dots,\mathbf{x}_N]$ and $\mathbf{T} = [\mathbf{t}_i,\dots,\mathbf{t}_N]$ are matrices formed by the samples $\mathbf{x}_i$ and $\mathbf{t}_i$ as columns, respectively. 

In our case, each sample $\mathbf{x}_i$ corresponds to an event, represented by a vector (with $D = 144$), as described in \hyperref[SS:ExperimentalProtocol]{Section} \ref{SS:ExperimentalProtocol}. For the three-class classification problems in our dataset, the elements of vectors $\mathbf{t}_i \in \mathbb{R}^C$ ($C = 3$ in our case) take values equal to $t_{ik} = 1$, if $\mathbf{x}_i$ belongs to class $k$, and, if $t_{ik} = -1$, otherwise. The solution of \hyperref[Eq:RRcriterion]{Eq.} \ref{Eq:RRcriterion} is given by:
\begin{equation}\label{Eq:RRsolution1}
\mathbf{W} = \mathbf{X} \left(\mathbf{X}^T \mathbf{X} + \lambda \mathbf{I} \right)^{-1} \mathbf{T}^T,
\end{equation}
or
\begin{equation}\label{Eq:RRsolution2}
\mathbf{W} = \left(\mathbf{X} \mathbf{X}^T + \lambda \mathbf{I} \right)^{-1} \mathbf{X} \mathbf{T}^T,
\end{equation}
where $\mathbf{I}$ is the identity matrix of appropriate dimensions. Here, we should note that, in our case, where the size of the data is big, $\mathbf{W}$ should be computed using \hyperref[Eq:RRsolution2]{Eq.} \ref{Eq:RRsolution2}, since the calculation of \hyperref[Eq:RRsolution1]{Eq.} \ref{Eq:RRsolution1} is computationally very expensive.

After the calculation of $\mathbf{W}$, a new (test) sample $\mathbf{x} \in \mathbb{R}^D$ is mapped on its corresponding representation in space $\mathbb{R}^C$, i.e. $\mathbf{o} = \mathbf{W}^T \mathbf{x}$, and is classified according to the maximal value of its projection, i.e.:
\begin{equation}
l_{\mathbf{x}} = \underset{k}{\arg max} \:\:o_k.
\end{equation}

\subsection{SLFN Network-based Nonlinear Regression}\label{SS:PVM}
\medskip

We also test the performance of a non-linear regression model. Since the application of kernel-based regression is computationally too intensive for the size of our data, we use a SLFN (\hyperref[fig:SLFN]{Fig.} \ref{fig:SLFN}) network-based regression model. Such a model is formed as follows:

\begin{figure}[h!]
\centering
\includegraphics[scale=0.30]{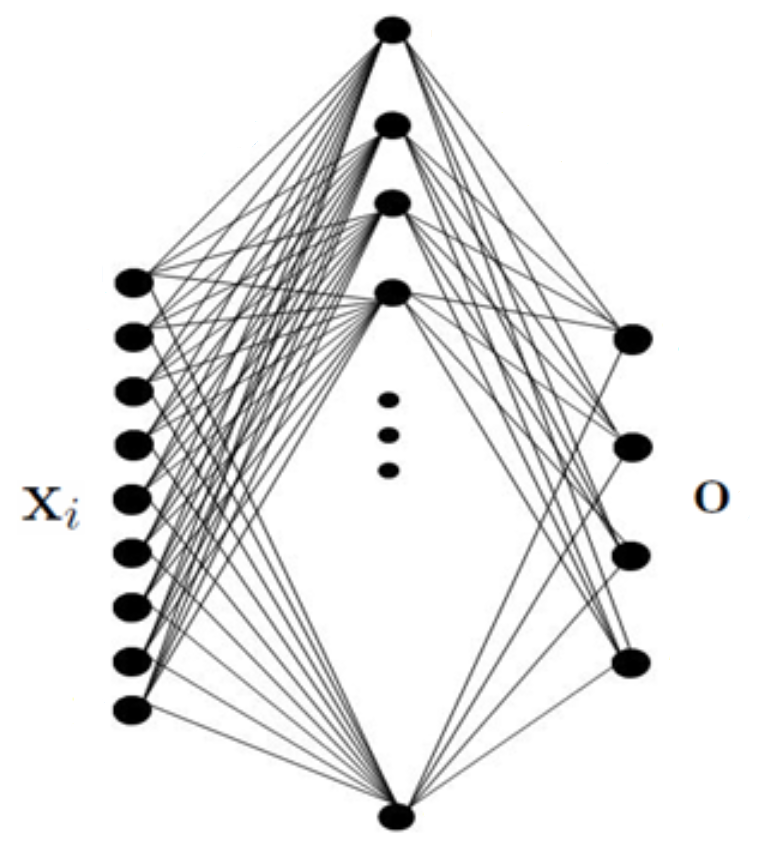}
\captionsetup{justification=centering}
\caption{SLFN}
\label{fig:SLFN}
\end{figure}

For fast network training, we train our network based on the algorithm proposed in \cite{huang2012extreme}, \cite{zhang2009prototype}, and \cite{iosifidis2017approximate}. This algorithm is formed by two processing steps. In the first step, the network's hidden layer weights are determined either randomly \cite{huang2012extreme} or by applying clustering on the training data. We apply $K$-means clustering in order to determine $K$ prototype vectors, which are subsequently used as the network's hidden layer weights.

Having determined the network's hidden layer weights $\mathbf{V} \in \mathbb{R}^{D \times K}$, the input data $\mathbf{x}_i, \:i=1,\dots,N$ are non-linearly mapped to vectors $\mathbf{h}_i \in \mathbb{R}^K$, expressing the data representations in the feature space determined by the network's hidden layer outputs $\mathbb{R}^K$. We use the radial basis function, i.e. $\mathbf{h}_i = \phi_{RBF}(\mathbf{x}_i)$, calculated in an element-wise manner, as follows:
\begin{equation}
h_{ik} = \exp{\left( \frac{ \|\mathbf{x}_i-\mathbf{v}_k\|^2_2 }{2 \sigma^2}  \right)}, \:\:k=1,\dots,K,
\end{equation}
where $\sigma$ is a hyper-parameter denoting the spread of the RBF neuron and $\mathbf{v}_k$ corresponds to the $k$-th column of $\mathbf{V}$. 

The network's output weights $\mathbf{W} \in \mathbb{R}^{K \times C}$ are subsequently determined by solving for:
\begin{equation}\label{Eq:PVMcriterion}
\mathbf{W}^* = \underset{\mathbf{W}}{\arg min} \:\: \| \mathbf{W}^T \mathbf{H} - \mathbf{T} \|^2_F + \lambda \|\mathbf{W}\|^2_F,
\end{equation}
where $\mathbf{H} = [\mathbf{h}_1,\dots,\mathbf{h}_N]$ is a matrix formed by the network's hidden layer outputs for the training data and $\mathbf{T}$ is a matrix formed by the network's target vectors $\mathbf{t}_i, \:i=1,\dots,N$ as defined in \hyperref[SS:RR]{Section} \ref{SS:RR}. The network's output weights are given by:
\begin{equation}\label{Eq:PVMsolution2}
\mathbf{W} = \left(\mathbf{H} \mathbf{H}^T + \lambda \mathbf{I} \right)^{-1} \mathbf{H} \mathbf{T}^T.
\end{equation}

After calculation of the network parameters $\mathbf{V}$ and $\mathbf{W}$, a new (test) sample $\mathbf{x} \in \mathbb{R}^D$ is mapped on its corresponding representations in spaces $\mathbb{R}^K$ and $\mathbb{R}^C$, i.e. $\mathbf{h} = \phi_{RBF}(\mathbf{x})$ and $\mathbf{o} = \mathbf{W}^T \mathbf{h}$, respectively. It is classified according to the maximal network output, i.e.:
\begin{equation}
l_{\mathbf{x}} = \underset{k}{\arg max} \:\:o_k.
\end{equation}

\section{Results}\label{SS:Results}
\medskip

In our first set of experiments, we have applied two supervised machine learning methods, as described in \hyperref[SS:RR]{Section} \ref{SS:RR} and \hyperref[SS:PVM]{Section} \ref{SS:PVM}, on a dataset that does not include the auction period. Results with the auction period will also be available. Since there is not a widely adopted experimental protocol for these datasets, we provide information for the five different label scenarios under the three normalization setups.
\medskip
\medskip
\begin{table}[h!]
\centering
\captionsetup{width=.816\textwidth} 
\caption[caption]{\\\hspace{\textwidth} Results Based on Unfiltered Representations}
\scalebox{0.84}{
\begin{tabular}{ccccc}
\hline
Labels & $RR_{Accuracy}$ & $RR_{Precision}$ & $RR_{Recall}$ & $RR_{F1}$  \\
\hline
1 	&	0,637 $\pm$ 0,055	& 0,505 $\pm$ 0,145	& 0,337 $\pm$ 0,003	& 0,268 $\pm$ 0,014	\\
2 	& 	0,555 $\pm$ 0,064	& 0,504 $\pm$ 0,131	& 0,376 $\pm$ 0,023	& 0,320 $\pm$ 0,050	\\
3 	& 	0,489 $\pm$ 0,061	& 0,423 $\pm$ 0,109	& 0,397 $\pm$ 0,031	& 0,356 $\pm$ 0,070	\\
5 	&	0,429 $\pm$ 0,049	& 0,402 $\pm$ 0,113	& 0,425 $\pm$ 0,038	& 0,400 $\pm$ 0,093	\\
10 	&	0,453 $\pm$ 0,054	& 0,400 $\pm$ 0,105	& 0,400 $\pm$ 0,030	& 0,347 $\pm$ 0,066	\\
\hline
Labels &  $SLFN_{Accuracy}$ & $SLFN_{Precision}$ & $SLFN_{Recall}$ & $SLFN_{F1}$ \\
\hline
1 	&	0,636 $\pm$ 0,055	& 0,299 $\pm$ 0,075	& 0,335 $\pm$ 0,002	& 0,262 $\pm$ 0,015	\\
2	& 	0,536 $\pm$ 0,069	& 0,387 $\pm$ 0,132	& 0,345 $\pm$ 0,009	& 0,260 $\pm$ 0,035 	\\
3 	& 	0,473 $\pm$ 0,074	& 0,334 $\pm$ 0,080	& 0,357 $\pm$ 0,005	& 0,270 $\pm$ 0,021	\\
5 	&	0,381 $\pm$ 0,038	& 0,342 $\pm$ 0,058	& 0,370 $\pm$ 0,020	& 0,327 $\pm$ 0,043	\\
10 	&	0,401 $\pm$ 0,039	& 0,284 $\pm$ 0,102	& 0,356 $\pm$ 0,020	& 0,290 $\pm$ 0,070	\\
\hline
\end{tabular}}
\label{tab:Raw}
\end{table}

\begin{table}[h!]
\centering
\captionsetup{width=.816\textwidth} 
\caption[caption]{\\\hspace{\textwidth} Results based on Z-score Normalization}
\scalebox{0.84}{
\begin{tabular}{ccccc}
\hline
Labels & $RR_{Accuracy}$ & $RR_{Precision}$ & $RR_{Recall}$ & $RR_{F1}$  \\
\hline
1	&	0,480 $\pm$ 0,040	& 0,418 $\pm$ 0,021	& 0,435 $\pm$ 0,029	& 0,410 $\pm$ 0,022	\\
2	& 	0,498 $\pm$ 0,052	& 0,444 $\pm$ 0,025	& 0,443 $\pm$ 0,031	& 0,440 $\pm$ 0,031	\\
3 	& 	0,463 $\pm$ 0,045	& 0,438 $\pm$ 0,027	& 0,437 $\pm$ 0,033	& 0,433 $\pm$ 0,034 	\\
5 	&	0,439 $\pm$ 0,042	& 0,436 $\pm$ 0,028	& 0,433 $\pm$ 0,028	& 0,427 $\pm$ 0,041	\\
10 	&	0,429 $\pm$ 0,046	& 0,429 $\pm$ 0,028	& 0,429 $\pm$ 0,043	& 0,416 $\pm$ 0,044	\\
\hline
Labels &  $SLFN_{Accuracy}$ & $SLFN_{Precision}$ & $SLFN_{Recall}$ & $SLFN_{F1}$ \\
\hline
1	&	0,643 $\pm$ 0,056	& 0,512 $\pm$ 0,037	& 0,366 $\pm$ 0,019	& 0,327 $\pm$ 0,046	\\
2	& 	0,556 $\pm$ 0,066	& 0,550 $\pm$ 0,029	& 0,378 $\pm$ 0,011		& 0,327 $\pm$ 0,030	\\
3 	& 	0,512 $\pm$ 0,069	& 0,497 $\pm$ 0,024	& 0,424 $\pm$ 0,047	& 0,389 $\pm$ 0,082	\\
5 	&	0,473 $\pm$ 0,036	& 0,468 $\pm$ 0,024	& 0,464 $\pm$ 0,028	& 0,459 $\pm$ 0,031	\\
10 	&	0,477 $\pm$ 0,048	& 0,453 $\pm$ 0,056	& 0,432 $\pm$ 0,025	& 0,410 $\pm$ 0,040	\\
\hline
\end{tabular}}
\label{tab:Z}
\end{table}

\begin{table}[ht!]
\centering
\captionsetup{width=.816\textwidth} 
\caption[caption]{\\\hspace{\textwidth} Results Based on Min-Max Normalization }
\scalebox{0.84}{
\begin{tabular}{ccccc}
\hline
Labels & $RR_{Accuracy}$ & $RR_{Precision}$ & $RR_{Recall}$ & $RR_{F1}$  \\
\hline
1	&	0,637 $\pm$ 0,054	& 0,499 $\pm$ 0,118	& 0,339 $\pm$ 0,005	& 0,272 $\pm$ 0,015	\\
2	& 	0,561 $\pm$ 0,063	& 0,467 $\pm$ 0,117		& 0,400 $\pm$ 0,028	& 0,368 $\pm$ 0,060	\\
3 	& 	0,492 $\pm$ 0,070	& 0,428 $\pm$ 0,111		& 0,400 $\pm$ 0,030	& 0,357 $\pm$ 0,072	\\
5 	&	0,437 $\pm$ 0,048	& 0,419 $\pm$ 0,078	& 0,429 $\pm$ 0,043	& 0,417 $\pm$ 0,063	\\
10 	&	0,452 $\pm$ 0,054	& 0,421 $\pm$ 0,110		& 0,399 $\pm$ 0,028	& 0,348 $\pm$ 0,066	\\
\hline
Labels &  $SLFN_{Accuracy}$ & $SLFN_{Precision}$ & $SLFN_{Recal}$ & $SLFN_{F1}$ \\
\hline
1	&	0,640 $\pm$ 0,055	& 0,488 $\pm$ 0,104	& 0,348 $\pm$ 0,007	& 0,291 $\pm$ 0,022	\\
2 	& 	0,558 $\pm$ 0,065	& 0,469 $\pm$ 0,066	& 0,399 $\pm$ 0,023	& 0,367 $\pm$ 0,050	\\
3 	& 	0,499 $\pm$ 0,063	& 0,447 $\pm$ 0,068	& 0,410 $\pm$ 0,032	& 0,370 $\pm$ 0,063	\\
5 	&	0,453 $\pm$ 0,038	& 0,441 $\pm$ 0,041	& 0,444 $\pm$ 0,030	& 0,432 $\pm$ 0,050	\\
10 	&	0,450 $\pm$ 0,048	& 0,432 $\pm$ 0,070	& 0,406 $\pm$ 0,037	& 0,377 $\pm$ 0,062	\\
\hline
\end{tabular}}
\label{tab:Min}
\end{table}

\medskip
\begin{table}[ht!]
\centering
\captionsetup{width=.816\textwidth} 
\caption[caption]{\\\hspace{\textwidth} Results Based on Decimal Precision Normalization}
\scalebox{0.84}{
\begin{tabular}{ccccc}
\hline
Labels & $RR_{Accuracy}$ & $RR_{Precision}$ & $RR_{Recall}$ & $RR_{F1}$  \\
\hline
1	&	0,638 $\pm$ 0,054	& 0,518 $\pm$ 0,132		& 0,341 $\pm$ 0,007	& 0,277 $\pm$ 0,018	\\
2 	& 	0,551 $\pm$ 0,066	& 0,473 $\pm$ 0,118		& 0,372 $\pm$ 0,018	& 0,315 $\pm$ 0,045	\\
3 	& 	0,490 $\pm$ 0,069	& 0,432 $\pm$ 0,113		& 0,386 $\pm$ 0,023	& 0,330 $\pm$ 0,059	\\
5 	&	0,435 $\pm$ 0,051	& 0,406 $\pm$ 0,115		& 0,430 $\pm$ 0,039	& 0,405 $\pm$ 0,095	\\
10 	&	0,451 $\pm$ 0,052	& 0,417 $\pm$ 0,108		& 0,399 $\pm$ 0,029	& 0,349 $\pm$ 0,067	\\
\hline
Labels &  $SLFN_{Accuracy}$ & $SLFN_{Precision}$ & $SLFN_{Recall}$ & $SLFN_{F1}$ \\
\hline
1	&	0,641 $\pm$ 0,055	& 0,512 $\pm$ 0,027	& 0,351 $\pm$ 0,007	& 0,297 $\pm$ 0,024	\\
2 	& 	0,565 $\pm$ 0,063	& 0,505 $\pm$ 0,020	& 0,410 $\pm$ 0,026	& 0,385 $\pm$ 0,054	\\
3 	& 	0,504 $\pm$ 0,061	& 0,465 $\pm$ 0,032	& 0,421 $\pm$ 0,040	& 0,393 $\pm$ 0,073	\\
5 	&	0,457 $\pm$ 0,038	& 0,451 $\pm$ 0,029	& 0,449 $\pm$ 0,031	& 0,438 $\pm$ 0,046	\\
10 	&	0,461 $\pm$ 0,053	& 0,453 $\pm$ 0,036	& 0,420 $\pm$ 0,035	& 0,399 $\pm$ 0,053	\\
\hline
\end{tabular}}
\label{tab:Dec}
\end{table}

The tables in this section provide details regarding the results of experiments conducted on raw data and three different normalization setups. We present these results, for our baseline models, in order to give insight into the pre-processing step for a dataset like ours, to examine the strength of the predictability of the projected time horizon, and to understand the implications of the suggested methods. Data normalization can significantly improve metric's performance in combination with the use of the right classifier. More specifically, we measure the predictability power of our models via the performance of the metrics of accuracy, precision, recall, and F1 score. For instance, \hyperref[tab:Raw]{Table} \ref{tab:Raw} presents the results based on raw data (i.e. no data decoding), and in the case of the linear classifier RR and label 5 (i.e. the $5^{th}$ mid-price event as predicted horizon), we achieve an F1 score of 40\%, where as in \hyperref[tab:Z]{Table} \ref{tab:Z} (i.e. the Z-score data decoding method), \hyperref[tab:Min]{Table} \ref{tab:Min} (i.e. min-max data decoding method), and \hyperref[tab:Dec]{Table} \ref{tab:Dec} (i.e. the decimal precision decoding method), we achieve 43\%, 42\%, and 40\%, respectively. This shows that in the case of the linear classifier, the suggested decoding methods did not offer any significant improvements, since the variability of the performance range is approximately 3\%. On the other hand, our non-linear classifier (i.e. SLFN) for the same projected time horizon (i.e. label 5) reacted more efficiently in the decoding process. SLFN achieves 33\% for the F1 score for non-normalized data, while the Z-score, min-max and decimal precision methods achieve 46\%, 43\%, and 43\%, respectively. As a result, normalization improves the F1 score performance by almost $10\%$. 

Normalization and model selection can also affect the predictability of mid-price movements over the projected time horizon. Very interesting results come to light if we try to compare the F1 performance over different time horizons.  For instance, we can see that regardless of the decoding method, the F1 score is always better for label 5 than 1, meaning that our models’ predictions are better further in the future. This result is significant, especially with unfiltered data and min-max and decimal precision normalizations, when F1 score is approximately 27\%, in the case of the one-step prediction problem (label 1), and 43\% in the case of the five-step problem (label 5).

Another aspect of the experimental results above stems from the pros and cons of linear and non-linear classifiers. More specifically, the RR linear classifier performed better on the raw dataset and for the Z-score decoding method in terms of F1 when compared to the SLFN (i.e. non-linear classifier). This is not the case for the last decoding methods (i.e. min-max and decimal precision), where our non-linear classifier presents similar or better results than RR. An explanation for this F1 performance discrepancy is due to each of these methods' engineering has. The RR classifier tends to be very efficient in high-dimensional problems, and these types of problems are linearly separable, in most cases. Another reason that RR can perform better when compared to a non-linear classifier is that RR can control the complexity by penalizing the bias, via cross-validation, using the ridge parameter. On the other hand, a non-linear classifier is prone to overfitting, which means that in some cases it offers a better degree of freedom for class separation.

\section{Conclusion}\label{SS:Conclusion}
\medskip

This paper described a new benchmark dataset formed by the NASDAQ ITCH feed data for five stocks for ten consecutive trading days. Data representations that were  exploited by order flow features were made available. We formulated five classification tasks based on mid-price movement predictions for 1, 2, 3, 5, and 10 predicted horizons. Baseline performances of two regression models were also provided in order to facilitate future research on the field. 
Despite the data size, we achieved an average out-of-sample performance (F1) of, approximately, $46\%$ for both methods. These very promising results show that machine learning can effectively predict mid-price movement.

Potential avenues of research that can benefit from exploiting the provided data include: a) prediction of the stability of the market, which is very important for liquidity providers (market makers) to make the spread, as well as for traders to increase liquidity provision (when markets can be predicted to be stable); b) prediction on market movements, which is important for expert systems used by speculative traders; c) identification of order book spoofing, i.e. situations where markets are manipulated by limit orders. While there is no spoofing activity information available for the provided data, the exploitation of such a large corpus of data can be used in order to identify patterns in stock markets that can be further analysed as normal or abnormal.

\section*{Acknowledgment}
\medskip

This work was supported by H2020 Project BigDataFinance MSCA-ITN-ETN 675044 (http://bigdatafinance.eu), Training for Big Data in Financial Research and Risk Management.


\bibliography{bibfile_copy.bib}

\end{document}